\documentstyle[12pt]{article}   
\begin{document}
\begin{titlepage}
\noindent
\begin{center}
{\bf FOCUSING ON THE FIXED POINT OF 4D SIMPLICIAL GRAVITY}\\
\vspace{1cm}
P. Bialas$^a$\footnote{Permanent address: Inst. Comp. Science, 
Jagellonian University, ul. Nawojki 11, 30-072 Krak\'ow, 
Poland.}, Z. Burda$^b\footnote{Permanent adress: Inst. of 
Physics, Jagellonian University, ul. Reymonta 4, 30-059 
Krak\'ow, Poland.}$, A. Krzywicki$^c$ and 
B. Petersson$^b$\footnote{bengt@phyaserv.uni-bielefeld.de}\\
\vspace{0.6cm}
$^a$ Inst. Theor. Fysica, Universiteit van Amsterdam, 1018 XE 
Amsterdam, The Netherlands\\
$^b$ Fakult\"at f\"ur Physik, Universit\"at Bielefeld, 
Postfach 10 01 31, D-33501 Bielefeld, Germany\\
$^c$ Laboratoire de Physique Th\'eorique et 
Hautes Energies, B\^{a}t. 211, Universit\'e de Paris-Sud, 
91405 Orsay, France\footnote{Laboratoire associ\'e au CNRS.}\\
\vspace{1cm}
{\bf Abstract}
\end{center}
Our earlier renormalization group analysis
of simplicial gravity 
is extended. 
A high statistics study of 
the volume and coupling constant 
dependence of the cumulants
of the node distribution 
is carried out. It appears that the phase
transition of the theory 
is of first order, contrary to what is generally 
believed .

\vfill
\par\noindent
January 1996
\par\noindent
LPTHE Orsay 96/08\\
BI-TP 96/05\\
ITFA-96-2
\end{titlepage}

\section*{1. Introduction}
The aim of this paper is to present results obtained
in a numerical study of the critical behavior of simplicial
gravity in four dimensions ($4d$). We limit our attention to
manifolds with the topology of a sphere, discretized 
according to the dynamical triangulation 
recipe \cite {dt}. The paper
has two facets: we extend our recent 
investigation \cite{bkk} of volume-volume correlations using the 
renormalization group (RG) techniques and we 
examine the finite-size scaling of the
integrals of 2- and 3-point 
curvature-curvature correlation functions. 
We attempt to achieve a better understanding of the theory 
putting together the two sets of data.
\par
A warning is in order before we proceed further. It is
obvious that we move in an almost uncharted teritory. Our
guide are analogies with more conventional field theory
and/or statistical mechanics and we are well aware of the
fact that these analogies might be misleading. Therefore,
we try to clearly separate the presentation of our data
from the speculations.
\par
The plan of the paper is the following: in sect. 2 we briefly
describe the lattice gravity model studied here as well
as the numerical algorithm employed in our computer simulations. 
Sect. 3 is devoted to the RG analysis. In sect. 4 the results
concerning the node distribution are presented. These results 
indicate, to our surprise, that the phase transition is 
presumably of first order. The last 
section contains the discussion
and conclusions. A summary of data 
is presented in tables grouped
in the Appendix. 

\section*{2. Model and numerical algorithm}
The model is identical to that studied in \cite{bkk}.
Thus, we take an Euclidean version of the Einstein-Hilbert
action, which for a $4d$ simplicial manifold reads
\begin{equation}
S= - \kappa_2 N_2 + \kappa_4 N_4
\label{action}
\end{equation}
\noindent
Here $N_k$ denotes the total number of $k$-simplexes. The
theory is defined by the partition function
\begin{equation}
Z(\kappa_2, \kappa_4) = \sum_{N_2 \: N_4} 
Z_{N_2 \, N_4} \, e^{-S}
\label{gcZ}
\end{equation}
\noindent
where
\begin{equation}
Z_{N_2 \, N_4} = \sum_{T(N_2, N_4)} W(T)
\label{mcZ}
\end{equation}
\noindent
The summation is over fixed topology $4d$ simplicial
manifolds $T(N_2, N_4)$ and $W(T)$ is the symmetry factor
taking care of the equivalent relabelings of a manifold.
\par
We are interested in a fixed $N_4$ canonical
ensemble of spherical manifolds with the partition function:
\begin{equation}
Z(\kappa_2, N_4)= \sum_{N_2} Z_{N_2 \, N_4} \, e^{-S}
\label{cZ}
\end{equation}
As in \cite{bkk} we actually simulate a grand-canonical
ensemble, but with a quadratic potential term 
${1 \over 2} \delta (N_4 - \bar{N_4})^2$ added to
 the action so that the value of $N_4$ at which the
 measurements are done is highly probable in the resulting
 $N_4$ distribution. This is the standard procedure. The
volume fluctuations hopefully
insure the ergodicity of the updating 
algorithm (cf. \cite{bbj}).
In most of the work we used $\delta=0.005$.
\par
The algorithm used in this paper comprises the five
standard local moves supplemented by the global ``baby
 universe surgery'' \cite{abbjp,aj}. The local moves
 are equiprobable. One local sweep is made up of $N_4$ 
attempts to perform a local move. ``Baby universe surgery'' 
consists in cutting out and gluing back, but at another
place, minimum neck baby universes (minBUs). All minBU
necks are identified but only minBUs with volume larger
than some lower cut-off are actually moved (the cut-off is set
for definiteness to 20 simplexes). After having identified
all necks the ``surgery'' is attempted a number of times equal
to twice the number of minBU necks. 
In a typical run nine sweeps of local moves are followed by 
one global sweep. The CPU time needed to perform a global
sweep depends on the number of baby universes and 
therefore on $\kappa_2$. In the neighbourhood 
of the critical point it is of the same order of magnitude as 
the time needed to perform a sweep of local moves.
\par
The global moves reduce the autocorrelation 
times dramatically
in the so-called cold phase, where the manifolds are highly
branched. They help much less elsewhere. 
In the vicinity of the
critical point minBU surgery 
reduces the autocorrelation time 
but does not prevent the critical 
slowing down to manifest itself 
vigorously. We show in Fig. 1
the time averaged integral of the autocorrelation function,
denoted by $\tau$, for a set of values of $\kappa_2$ and $N_4$.
We have measured $N_0$ once every 10 sweeps 
\footnote{We 
have checked that at the phase transition 
point the frequency of  passes 
through $N_4 = \bar{N_4}$ is the same
in both phases.} (local and global sweeps are 
counted on the same footing). 

\section*{3. Monte Carlo Renormalization Group}
We follow the approach proposed in \cite{bkk} based on the 
hierarchical baby-universe structure of a 
typical dynamically 
triangulated manifold. The RG 
transformation, appropriately called
{\em fractal blocking} in ref. 
\cite{ct}, consists in cutting
the outer layer of minBUs.
\par
Let $\mid x_A - x_B \mid$ denote 
the geodesic distance between
simplexes $A$ and $B$ (i.e. the minimum 
number of steps, along the
dual lattice, separating $A$ 
and $B$). The correlation function
\begin{equation}
G(r) = \langle N_4^{-2} \sum_{A B} 
\delta(r - \mid x_A - x_B \mid) \rangle
\label{corr1}
\end{equation}
characterizes the global geometry 
of the manifold. Its moments
$\langle r^k \rangle$, $k=1,2,...$, 
are dimensionless lattice
observables which transform under RG as
$\langle r_{in} \rangle \to \langle r_{out} \rangle 
= \langle r_{in} \rangle - \delta r$, etc. 
\par
Our RG transformation can be 
regarded as corresponding to the
increase of the lattice spacing $a$. 
Indeed, let us keep fixed 
the physical volume $N_4 a^4$ of the manifold. Then
\begin{equation}
\delta \ln{{1 \over a}} = {1 \over 4} \delta \ln{N_4}
\label{step}
\end{equation}
\noindent
This change of the resolution should be accompanied by a
shift of coupling constants. Assuming that $\kappa_2$ is
the only coupling that matters for the global shape of the
manifold one writes
\begin{equation}
\delta r = r_N \delta \ln{N_4} + r_\kappa \delta \kappa_2
\label{rgeq}
\end{equation}
\noindent
where $r_N$ and $r_\kappa$ denote partial derivatives
 of $\langle r \rangle$
with respect to $\ln{N_4}$ and $\kappa_2$. Eq.
(\ref{rgeq}) is used to
find $\delta \kappa_2$ and, together with
(\ref{step}) to estimate the $\beta$-function
defined by
\begin{equation}
\beta (\kappa_2) = {{\partial \kappa_2} \over 
{\partial \ln{{1 \over a}}}}
\label{beta}
\end{equation}
\noindent
We have calculated the $\beta$-function for 
$N_4=$ 4000, 8000 and 16000. The results 
are shown in Fig. 2. 
\par
Of course, one could use another moment of $G(r)$.
If our philosophy is correct the resulting 
$\beta$-function should be the
same as before, at least in the limit of very 
large $N_4$. We have measured
$\langle r^2 \rangle$ along with
$\langle r \rangle$. The $\beta$-function
obtained using $\sqrt{\langle r^2 \rangle}$ 
is almost undistinguishable from that
calculated from $\langle r \rangle$.
\par
We have also computed the ratio
\begin{equation}
B  = {{\sqrt{\langle r_{in}^2 
\rangle}/\langle r_{in} \rangle} \over 
{\sqrt{\langle r_{out}^2 \rangle}/\langle r_{out} \rangle}},
\label{scrat}
\end{equation}
\noindent
which is found to be very close to unity for all values 
of $\kappa_2$ and $N_4$ considered in this paper. 
The best fit to
$B =$ const yields 1.0014(56) , 0.9993(54) and 1.0009(31) 
for $N_4=$ 4000, 8000 and 16000
respectively. Thus, the shape of $G(r)$ is approximately
constant along the RG flow provided that $r$ is measured in units
of $\langle r \rangle$ (compare with \cite{dbs}).
\par
These results are nicely self-consistent and confirm
the results presented in \cite{bkk}. The zero of the
$\beta$-function is close to the position of the critical point
determined by different methods. 
It moves up in $\kappa_2$
as $N_4$ increases but seems to converge to a limiting value.
\par
The slope of the $\beta$-function does not vary significantly
with $N_4$. The average value obtained combining data at 
$N_4=$ 4000, 8000 and 16000 is 
$\beta_0 \equiv -\beta'(\kappa_2^*) = 7.32(46)$.
This value should be for the moment 
taken with circumspection: we
do not know how reliable is our blocking scheme. 
It would be interesting to calculate this 
parameter using another RG blocking scheme, 
for example one derived from the proposal set forth 
in Ref. \cite{ct}.
\par
As already 
explained in \cite{bkk} our blocking
operation cannot be iterated. This would 
require being able to cut baby
universes with necks larger than the minimum one, 
which is very difficult to
implement. Thus the standard method of 
estimating finite-size corrections,
which consists in comparing results at 
the same lattice volume but 
after different number of blocking steps, 
cannot be applied here. All 
we can do is to compare data obtained at different volumes.  
\par
Consider now
the intrinsic Hausdorff dimension $d_H$. It has been observed
in refs. \cite{aj,dbs,ckr,abjk} that $d_H$ is 
very large, perhaps infinite,
in the hot phase and close to 2 in the cold one. 
Indeed (and this can be also seen in the tables 
presented in the Appendix) $\langle r \rangle$ depends 
weakly on $N_4$ below the critical point and rises roughly like
$N_4^{1/2}$ above it. This result concerns $d_H$ computed
at {\em fixed} $\kappa_2$. However, as already mentioned 
in \cite{bkk}, it is rather the value of $d_H$ attached 
to a RG trajectory which has a real physical significance.
Unfortunalely, our calculation of the $\beta$-function 
does not extend to large enough lattices and is not 
sufficiently precise to make possible 
a serious determination of RG trajectories.
We can, however, use the obvious equation

\begin{equation}
d_H = \langle r \rangle \delta \ln{N_4}/\delta r
\label{Haus}
\end{equation}

\noindent
to estimate the (effective) fractal dimension locally, 
on a RG trajectory passing through a given point in 
the $(N_4, \kappa_2)$ plane. Points close to $\kappa_2^*$
are of particular interest. Using the RG data given in 
the Appendix we find at $N_4 = 8000$ the fractal 
dimensions $3.7(1.4)$ and $4.4(1.3)$ for $\kappa_2 = 1.200$ 
and $1.225$, respectively. At $N_4 = 16000$ we obtain 
$d_H = 4.0(1.4), 5.3(2.6)$ and $5.7(2.9)$ for $\kappa_2 = 
1.225, 1.238$ and $1.250$, respectively (see \cite{dbs} for
a similar result).

\section*{4. Finite-size scaling analysis}
Our next set of data concerns the (normalized) cumulants of
the distribution of the number $N_0$ of nodes of the lattice:

\begin{equation}
\begin{array}{ccl}
c_2(N_4)& =&{1 \over N_4} [ \langle N_0^2 \rangle -
 \langle N_0 \rangle^2 ] \\& & \\
c_3(N_4)& = &{1 \over N_4} [\langle N_0^3 \rangle - 
3 \langle N_0^2 \rangle
\langle N_0 \rangle + 2 \langle N_0 \rangle^3 ]
\end{array}
\label{cums}
\end{equation}

\noindent
The average is computed in the canonical ensemble. 
For a manifold with spherical topology and $N_4$
fixed, the number of nodes $N_0$ is linearly
releted to $N_2$~:
$N_0 = N_2/2 - N_4 + 2$. Therefore the cumulants 
correspond to the second and the third derivative 
of the free energy with respect to $\kappa_2$
or, in other words, to the lattice version of the 
integrated 2- and 3-point connected correlators 
of the operator $\sqrt{g}R(x)$, where $R(x)$ denotes the
scalar curvature. The two-point connected
curvature--curvature correlator can be defined in 
an invariant manner, for example, as

\begin{equation}
c_2(r, V) = V^{-1} \Big\langle \int d^4x \sqrt{g(x)}d^4y 
\sqrt{g(y)} \;  \delta(r - \mid x-y \mid) [R(x)R(y)- 
\bar{R}^2] \Big\rangle ,
\label{correlator}
\end{equation}

\noindent
where $\mid x-y \mid$ is the geodesic distance between 
the points $x$ and $y$, $V$ is the (fixed) volume and 
$\bar{R}= V^{-1}\langle \int d^4x \sqrt{g(x)} R(x) 
\rangle$. We use here the more transparent continuum notation 
for convenience. Integrating $c_2(r,V)$ over $r$ one gets the
heat capacity $c_2(V) = V\big\{ \langle \bar{R}^2 \rangle
- \langle \bar{R} \rangle \langle \bar{R} \rangle \big\}$.
Notice that putting instead of the square bracket 
the expression $[R(x)-\bar{R}][R(y)-\bar{R}]$ 
one obtains a distinct
correlator, but with the same integral. 
The non-uniqueness of the
curvature-curvature correlator has 
already been noticed in \cite{dbs2}.
For our purposes it is not necessary 
to settle this problem. We merely 
assume that there exists a definition 
compatible with the finite-size 
scaling of the integrated correlator.
\par
In standard statistical mechanics the $r$-dependence of the 
two-point correlator in the 
neighbourhood of the continuous phase 
transition point is

\begin{equation}
c_2(r) \sim r^{z}  C(\mid \kappa_2-\kappa_2^* \mid^\nu r) 
\label{cf}
\end{equation}

\noindent
with $C(x)$ falling faster than a polynomial. The
fall off of the correlator is controlled by the value of the mass
gap $\sim |\kappa_2 - \kappa_2^*|^\nu$. The inverse of the mass 
gap, i.e. the correlation lentgth, is the only relevant length 
scale. This has as a consequence the following finite-size 
scaling~:

\begin{equation}
\begin{array}{ccl}
c_2(N_4) & = & N_4^b f[(\kappa_2 - \kappa_2^*)N_4^c] \\ & & \\
c_3(N_4) & = & N_4^{b+c} f'[(\kappa_2 - \kappa_2^*) N_4^c]
\end{array}
\label{scal}
\end{equation}

\noindent
A pedestrian derivation of (\ref{scal}) consists 
in integrating both sides of (\ref{cf}) over a region 
of finite linear extension $\sim N_4^{1 \over d_H}$, $d_H$ 
being the internal Hausdorff dimension of the system, 
at the critical point. 
One then gets (\ref{scal}) with $b=\alpha/d_H\nu$ 
and $c=1/d_H \nu$, where $\alpha = \nu (1+z)$ is to be 
identified with the specific heat exponent. Since 
$\alpha < 1$ for a continuous transition, one gets $b,c < 1$ 
using Fischer's hyperscaling relation $\alpha = 2 - d_H\nu$. 
\par
The equations (\ref{scal}) also hold when
the transition is of first order. 
In this case one most easily obtain (\ref{scal}) using 
the fact that the distribution of
the internal energy of the system has a bimodal shape in the
neigborhood of the critical point. The finite-size scaling 
then reflects the existence of tunneling between the 
two maxima. One easily finds that the exponents $b$ 
and $c$ are now strictly equal to unity: $b=c=1$.
\par
The function $f(x)$ in (\ref{scal}) is bell-shaped. 
Three values of its
argument are of particular interest: they correspond
to the maximum, zero and minimum of the derivative $f'(x)$.
In what follows they are referred to by using the appropriate 
subscripts. Thus $f'(x_{max}) = \max{[f'(x)]}$,
$f'(x_{min}) = \min{[f'(x)]}$, $f'(x_0) = 0$ and, 
of course, $f(x_0) = \max{[f(x)]}$. 
\par
We read the 
scaling of the maximum of the second 
cumulant with $N_4$ from the first of eqs. (\ref{scal})~:

\begin{equation}
\max{[c_2(\kappa_2)]} \sim N_4^b 
\label{power}
\end{equation}

\noindent
the proportionality constant being equal to $f(x_0)$. 
The location of this maximum is $\kappa_2 = \kappa_{2 \, 0}$ 
and is read from the equation  $x=x_0$~:

\begin{equation}
\kappa_{2 \, 0} = \kappa_2^* + \frac{x_0}{N_4^c}
\label{posi}
\end{equation}

\noindent
Analogously one finds

\begin{equation}
\begin{array}{ccl}
\max{[c_3(\kappa_2)]}& \sim &N_4^{b+c}\\& &\\
\min{[c_3(\kappa_2)]} &\sim &N_4^{b+c}
\end{array}
\label{power3}
\end{equation}

\noindent
The location of these structures is found from the right-hand
side of eq. (\ref{posi}) after replacing $x_0$ by $x_{max}$ and
$x_{min}$, respectively.
\par
The finite-size scaling functions in (\ref{scal}) do not
include contributions to the cumulants originating from 
the non-singular terms in the free energy. This background
must however be taken care of in the data analysis. 
Furthermore, the scaling sets in only at large 
enough volumes and, consequently, 
there are sub-leading corrections to the powers of 
volume appearing in the scaling formulae themselves.
Thus analysing data taken at not too large $N_4$ 
one should, in general, replace $N_4^c$
appearing in the argument of the scaling function by 
$N_4^c ( 1 + g N_4^\omega)$, say. 
\par
After all these preliminaries we are in a position to
discuss the data produced in our numerical experiment
(compare with \cite{aj0,ckr}). The second and the third 
cumulant, calculated for volumes 
ranging from $N_4=4000$ to 32000
are shown in Figs. 3 and 4, respectively,
as functions of the coupling $\kappa_2$. The curves and 
the error window have been obtained using 
reweighting \cite{fs} together with the
jack-knife method.
The reweighting is a clever interpolation
procedure, which combines in a non-trivial manner 
data obtained from runs with
different coupling. This corresponds to an 
effective increase of the statistics and therefore 
the error window of the resulting curves tends to be
smaller that the size of the errors
of single measurements. However, reweighting is truly
efficient when the set of couplings is sufficiently dense.
Otherwise, although the procedure is doing 
its best to interpolate properly,
the curve can occasionally miss some points. 
We have been more interested in the heights of peaks than
in the exact shape of the curves. Therefore, we have decided to
use our computer time to make precise measurements at these
special points, where we had to fight against the critical
slowing down, instead of making less precise measurements but on a
dense grid of points.
\par
Before entering the detailed quantitative discussion of these
data, let us see what can be concluded from a rapid perusal
of the figures. First of all, the curves corresponding to
different volumes look very similar, modulo rescaling of the 
coordinate axes. Such a behavior is precisely what finite-size
scaling predicts. One also notices that as one moves away from
the critical point the cumulants come down to a higher value
on the left than on the right-hand side. We shall see later
hat this asymmetry is to large extent due to a 
$\kappa_2$-dependent background.
\par
A more careful look at our figures 
indicates that the rate 
of growth of the structures 
increases with the volume. In
particular, the change occuring as 
one moves from $N_4=16000$ to
$N_4=32000$ is striking. As soon as 
we could contemplate the cumulant
data at $N_4=32000$ we were led to 
suspect that a change of regime
does occur in this last volume interval. 
Consequently we increased
our statistics at $N_4=32000$ and 
close to the critical point, in
order to see fine structures in the 
data. The result of this effort 
is shown in Fig. 5, where we display the energy ($N_0$)
histogram with two clearly separated peaks, obtained at 
$N_4=32000$ and $\kappa_2=1.258$ . The run history  
shows the system wandering between 
the two states. As one shifts
$\kappa_2$ to 1.252 (or 1.264) 
the bimodality of the histogram
stops being visible, at least with 
our statistics. However, the
histogram is skew, compared to 
histograms observed at values of
$\kappa_2$ more distant from the 
critical one or corresponding
to smaller volumes. Furthermore, 
the run history shows 
that the system makes recurrent 
excursions to a neighbor state,
where it does not stay for long, however. 
\par
Let us summarize what has been learned so far: all our data
at $N_4 \leq 16000$ are compatible with the transition being
continuous. But a double peak structure 
in the energy distribution is
observed at $N_4=32000$. Such a bimodality developing with  
increasing volume is a signal of a first-order transition. 
The increase of the rate of growth of the structures in the
critical region points in the same direction. In the rest of
this section we show that the most natural description
of all our data is obtained assuming that the exponents $b$
and $c$ are indeed equal to unity.
\par
The estimated values of $\max{[c_2]}$ can be found
in the last table of the Appendix. 
Assuming $\max{[c_2]}$ grows like in (\ref{power}), i.e.
as a power of the volume, one can calculate the exponent
for successive pairs of neighbor volumes. One finds
$0.30(5)$, $0.42(7)$ and $0.81(9)$ 
for the first, second and third pair of points. Thus
the effective exponent increases with the volume.
When a fit to all points is attempted
with the power low (\ref{power}) the 
exponent is $b=0.41(2)$.
However the fit is bad, $\chi_2/dof = 23.5$, and it misses the
point at $N_4=32000$.
\par
A considerably better description 
is obtained from a linear fit,
assuming a nonvanishing background. 
Assuming that $\max{[c_2]}$ is linear in both
$N_4$ and $\kappa_{2 \, max}$  we obtain the fit 
$\max{[c_2]} = 0.80(7) \times 10^{-5} N_4 + 0.267(94)
 - 0.161(90) \kappa_{2 \, max}$, with $\chi^2/dof = 4.6$. 
The background, which corresponds to the second
and third term in this fit, is shown as a dotted straight line 
in Fig. 3. Notice that a background decreasing with $\kappa_2$
fits well with the observed structure of the histogram in 
Fig. 5, where the left peak, corresponding 
to the lower $\kappa_2$
phase, is wider than the right one.  
\par
Consider now the third cumulant. A quantity of particular
interest is $\max{[c_3]} - \min{[c_3]}$, since it is 
independent of the $\kappa_2$-independent part of the
background. We fit its $N_4$-dependence with 
a fixed power and with
a parabola. In the former case we get
an exponent $1.79(14)$ close
to the first-order transition exponent $2$ ($\chi^2/dof = 10.0$). 
Assuming that the exponent is 2 but introducing 
next-to-leading corrections we fit better the data with the
parabola $0.30(4) \times 10^{-7} N_4^2 - 
0.17(9) \times 10^{-3} N_4 + 
0.9(3)$ ($\chi^2/dof = 5.5$).
\par
The last part of the data analysis concerns the positions
of the characteristic points of the scaling function $f(x)$. 
We first try the formula (compare with (\ref{posi})) 
 
\begin{equation}
\kappa_{2 \, j}(N_4) = \kappa_2^* + \frac{x_j}{N_4^c}
\label{fit1}
\end{equation}

\noindent
with $j=\{ max, 0, min \}$ to fit $12$ 
data points corresponding to 
four volumes and three positions: 
maximum, zero and minimum of the 
third cumulant.  In the functions 
$\kappa_{2 \, j}(N_4)$ we keep the values of
$\kappa_2^*$ and $c$ common, so that the three curves
only differ by the numerator $x_j$ of the second term.
Hence, altogether we have $12$ data points and five 
free parameters.  We obtain $c=0.47(7)$, 
$\kappa_2^*=1.327(13)$ and $\chi^2/dof = 24.5$.
\par
An alternative fit, more appropriate 
for a first-order transition,
is obtained using the formula

\begin{equation}
\kappa_{2 \, j}(N_4) = \kappa_2^* + \frac{x_j}{N_4+a_4}
\label{fit2}
\end{equation}

\noindent
The number of free parameters is the same as before.
We get now $\kappa_2^* = 1.293(4)$, $a_4 = 5234(1018)$ and
$\chi^2/dof = 8.3$. The other parameters are 
$x_{min} = -1.54(17) \times 10^3, x_0 = -1.39(17) \times 10^3, 
x_{max} = -1.16(15) \times 10^3$. 
The fit is illustrated in Fig. 6, where $\kappa_{2 \, j}(N_4)$
should be read on the vertical axis 
and $1000/(N_4 + 5234)$ on
the horizontal one. 
\par
Our conclusions rest heavily on results obtained at
$N_4=32000$ and are sensitive to data taken at small $N_4$. 
Therefore, it will be very important to extend
this study to larger lattices. Until this is done our
conclusions will remain somewhat tentative. Extrapolating
the last fit we expect that the bimodality of the $N_0$ 
histogram will show up at in the neighborhood of 
$\kappa_2 = 1.267$ for $N_4 = 48000$ 
and at $\kappa_2 = 1.273$ for $N_4 = 64000$\footnote{While 
completing our manuscript we have submitted 
a job with $N_4 = 48000$, 
$\kappa_2=1.267$.  A flip between the phases occurs 
after roughly $20000$ sweeps. B.V. de Bakker and 
J. Smit being informed about our predictions have 
looked at their data files
in search for the expected double peak structure and
have indeed found a time history flipping between two 
states for $N_4 = 64000$ near the expected value 
of $\kappa_2$.}.

\section*{5. Discussion}
  The most important new result of this paper is
the evidence that the phase transition of the
model is of first order. We have started this
study persuaded, as everybody, that in $4d$ the 
transition is continuous. The surprising
discovery of signals of a discontinuous transition
requires a reevaluation of the standard picture.
Therefore it is important to formulate a number of
caveats.
\par  
   One can, of course, 
extend the study to larger systems in order to check
the expectations formulated in the preceding section.
Most likely these expectations will be confirmed if
a standard simulation set-up is used. One can wonder, 
however, whether or not the discontinuous 
phase transition is 
an artifact of the algorithm employed. The latent heat
of the transition is quite small (e.g. in 
comparison with $3d$, where a first order transition
has been observed). The moves employed in the simulation
are not suspect by themselves. However, the constraint
limiting volume fluctuations might not be innocuous 
\footnote{This has been already mentioned by other people,
in particular in ref. \cite{ctkr}, where an attempt has 
been made to check the point at $\kappa_2 = 0$. No signal
of a breakdown of ergodicity was found, but the accepted
deviations from the reference volume were still  
modest.}. Strictly speaking, the proof of ergodicity does
hold when there is no limitation imposed on multiplicity
of simplexes in intermediate configurations. 
It might be that there exists a set of paths 
requiring large departures from the volume where
the measures are performed and connecting the two states 
we observe at the transition point. Then, by excluding 
these paths one would create artificially a "potential 
barrier". We have checked that our results are not 
affected when one changes the parameter $\delta$, 
multiplying the quadratic term in the action, by an 
order of magnitude (from 0.005 to 0.0005). Further 
investigation in this direction would be important
and desirable.
\par
Our renormalization group study confirms and 
strengthens the results presented in ref. \cite{bkk}.
There is a value of $\kappa_2$, close to the 
transition point found by other 
methods, where the manifolds appear to show 
a self-similar structure under the RG transformation. 
In the neighbourhood of this point the intrinsic 
Hausdorff dimension, measured along the 
RG flow, is close to $d_H=4$ instead of being 2 or $\infty$.
Thus, from these data alone one would claim that 
the model has an isolated ultra-violet 
stable fixed point and that one 
can define a sensible continuum limit 
letting $\kappa_2 \to \kappa_2^*, \; N_4 \to \infty$ 
in a way insuring that the lattice spacing $a$ tends to
zero when measured in physical units
\footnote{This limit would presumably preserve the global
geometry of manifolds, in particular the fractal dimension $d_H$.
The existence of the graviton is a separate issue. Notice
that in $2d$, where the graviton is absent,  one can 
nevertheless define a continuum limit.} .
The RG analysis has been performed for volumes 
up to $N_4=16000$ while the bimodal distribution of $N_0$ 
is only observed at $N_4=32000$. If the first-order phase transition 
signal is an algorithm artifact and if 
the algoritm becomes unreliable only at 
volumes larger than $N_4=16000$ then our
RG results may still retain their full significance.
\par
  If the transition is not an artifact 
then the relevance of our RG study is
uncertain. This failure of the RG method to 
show a discontinuous transition could be
attributed to either an
inadequacy of the blocking method 
or to the presence of finite size
effects which, as we have discussed, 
cannot be eliminated in the present
formulation.
\par
It should be stressed that our RG results are based on the
volume-volume correlations only. The corresponding 
characteristic length scale is $\langle r \rangle$. 
To further understand the 
dynamics of the model it is necessary to study also the 
curvature-curvature correlations, which are controlled 
by another length scale, essentially the inverse mass 
of the lattice "graviton". This has been the motivation 
of our finite-size scaling analysis of the moments
of the node distribution.
\par
  It is not surprising to have a finite mass gap on 
the lattice, i.e. as long as one 
does not take the continuum
limit. Indeed, the discrete theory is not endowed with 
the continuous gauge symmetry implying the existence 
of the graviton. But it is natural 
to expect that this mass gap
scales to zero as one approaches the critical point,
so that in the continuum limit one has a chance of 
recovering a massless graviton. But this 
means that the transition is continuous. If this reasoning
is correct and if the discontinuity of the 
phase transition is an intrinsic feature of 
the model than the latter is not an 
acceptable model of quantum gravity.
 \footnote{It has been pointed out to us by 
J. Smit that the continuity of the 
transition is perhaps not 
necessary: There exist models 
with a Coulomb phase and no continuous
transition. In the $Z(N)$ gauge model studied in 
ref. \cite{ab} the Coulomb phase appears at finite $N$.
Due to quantum tunneling the model anticipates the $U(1)$
symmetry which becomes manifest in the action in the limit
$N \to \infty$ only.}
\par
   It is clear that the surprising discovery
presented in this paper raises a series of questions.
Let us hope that they will be rapidly resolved by 
a joint effort of all groups interested in the 
development of lattice gravity. 

\section*{Acknowledgements}
We thank B.V. de Bakker, J. Jurkiewicz and J. Smit
for conversations. We are indebted to J. Jurkiewicz 
for providing us the opportunity to check our code
against his. We are grateful to B. Klosowicz 
for very precious help. We are indebted to the CNRS 
computing center IDRIS, and personally to V. Alessandrini, 
for computer time and cooperation. We are also indebted 
to IC3A for computer time on {\sc PowerXplorer} at SARA 
and the HLRZ J\"{u}lich for computer time on PARAGON.
P.B. thanks the Stichting voor Fundamenteel Onderzoek 
der Materie (FOM) for financial support. Z.B. has 
benefited from the financial support of the
Deutsche Forschungsgemeinschaft under the contract 
Pe 340/3-3.

\newpage
\section*{Figure captions}
\begin{description}
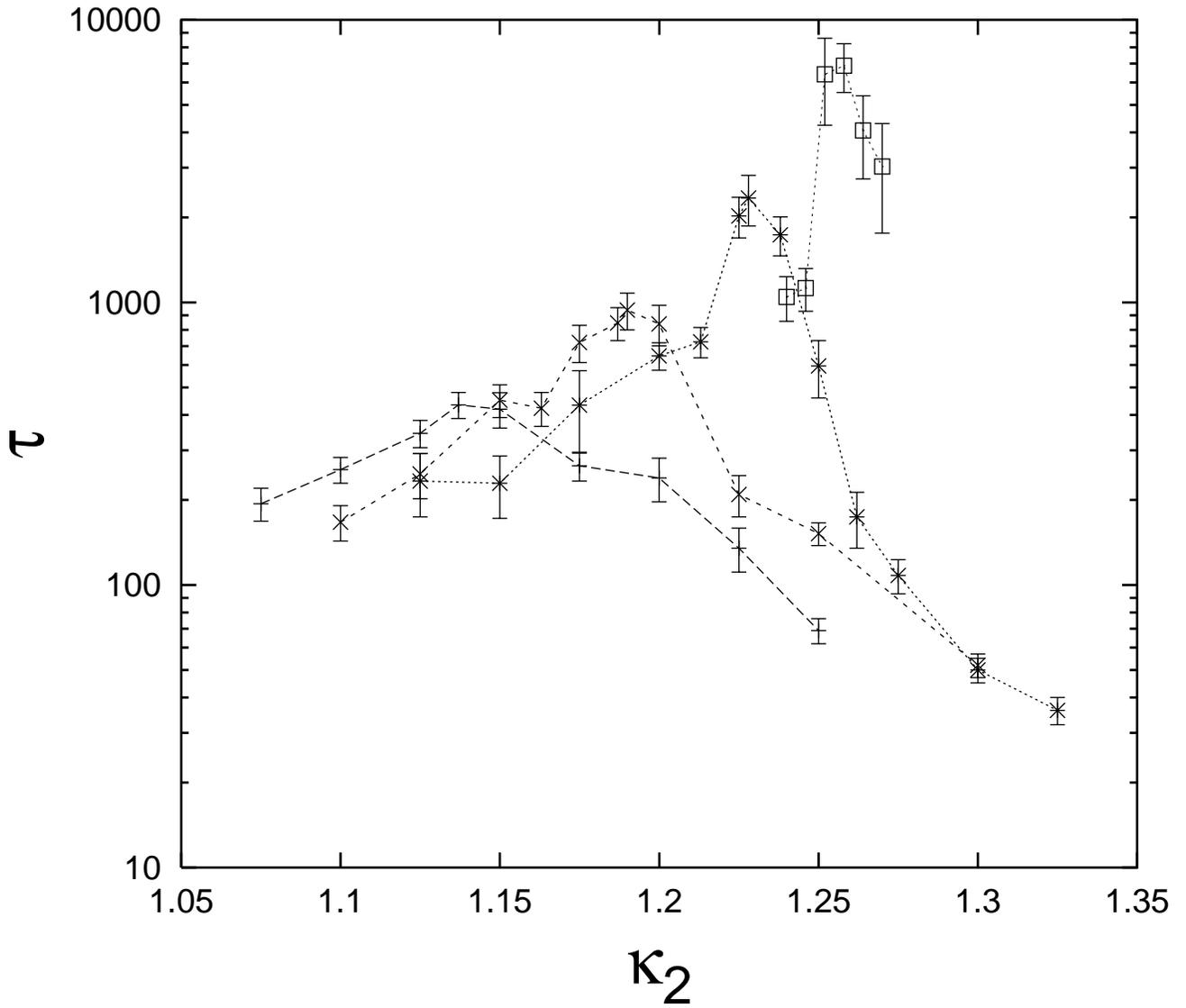
\item[Fig. 1] - The integral $\tau$ of the autocorrelation 
function (in sweeps) versus $\kappa_2$ for different volumes. 
The symbols corresponding to different volumes are
$4000(+)$, $8000(\times \hspace{-0.32cm} -)$, 
$16000(\times)$ and $32000(\Box)$.
The quantity measured 
is $N_0$. The $\tau$'s for other observables are comparable,
with the exception of $N_{4out}$, for which $\tau$ is never
larger than 100 sweeps.

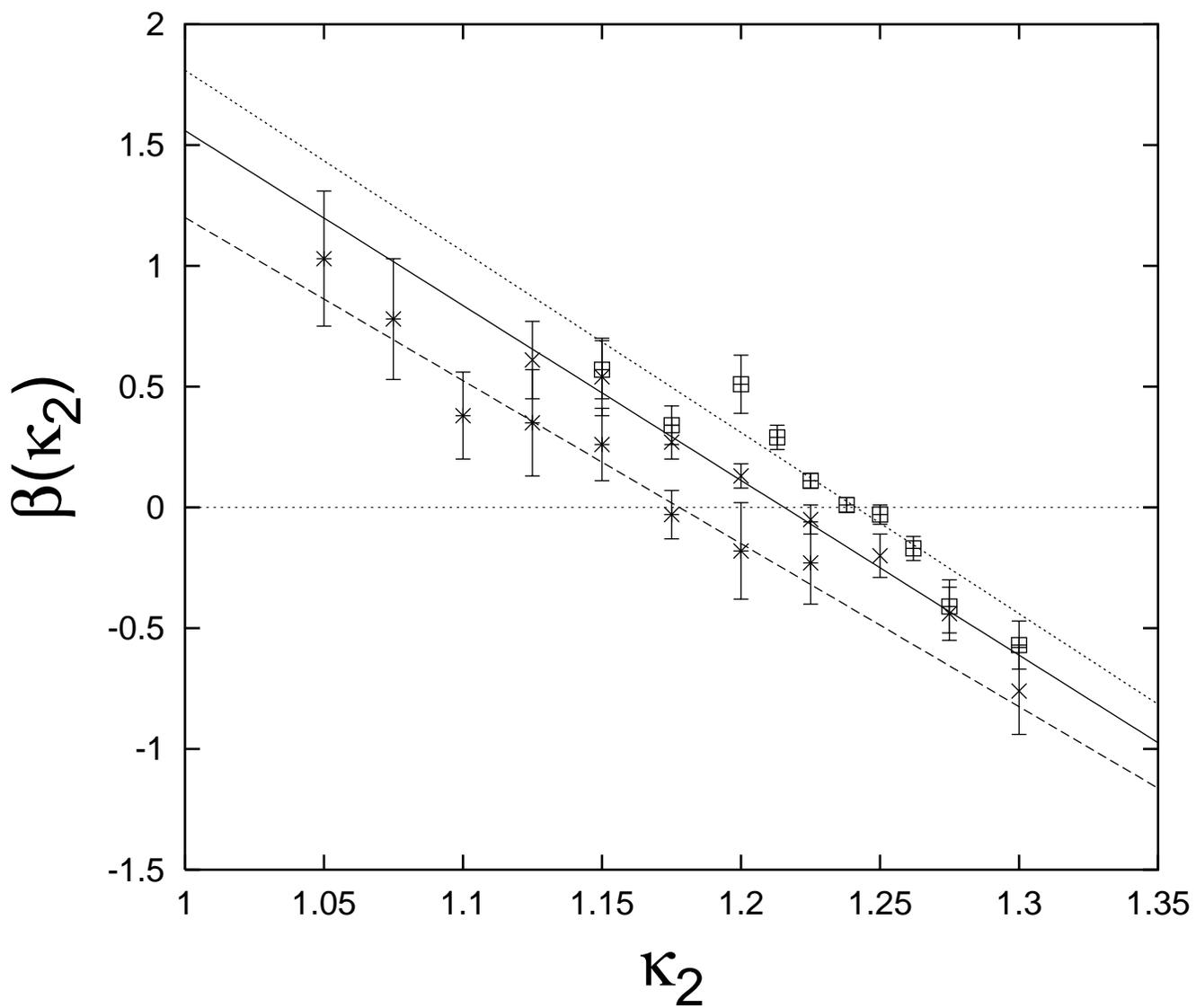
\item[Fig. 2] - The nonperturbative $\beta$-function for 
$N_4 = 4000(\times \hspace{-0.32cm} -)$, $8000(\times)$ and $16000(\Box)$. 
The zeros are at $\kappa_2 = 1.178(10), 1.217(4)$ and $1.240(2)$, 
respectively.

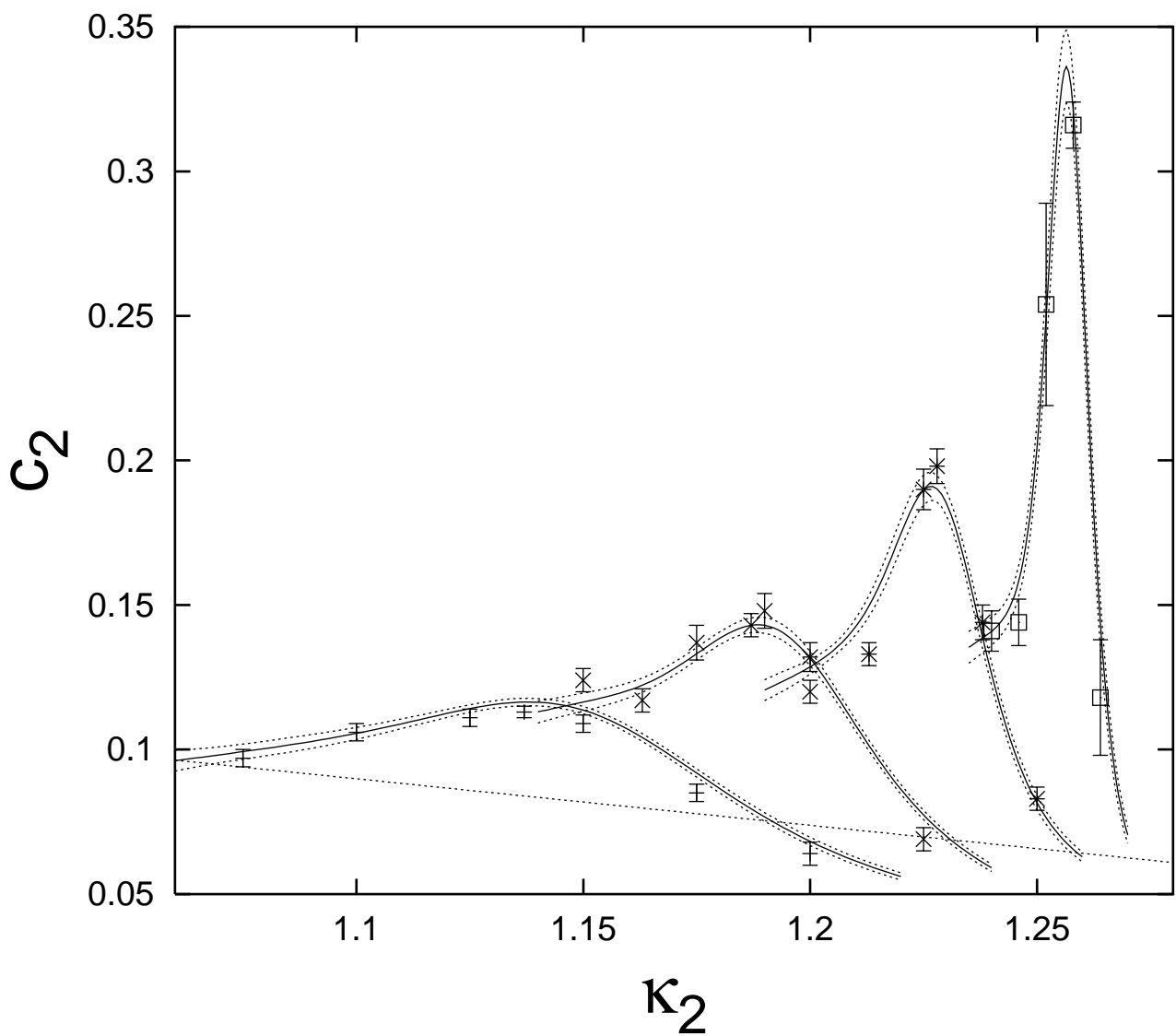
\item[Fig. 3] - The second cumulant $c_2$ as a function 
of $\kappa_2$ for lattice size 
$N_4 = 4000(+)$, $8000(\times)$, 
$16000(\times \hspace{-0.32cm} -)$ and $32000(\Box)$. 
The distance between the upper
and lower lines gives the error window.
The dotted line is a background fitted
to the linear form as discussed in the text.

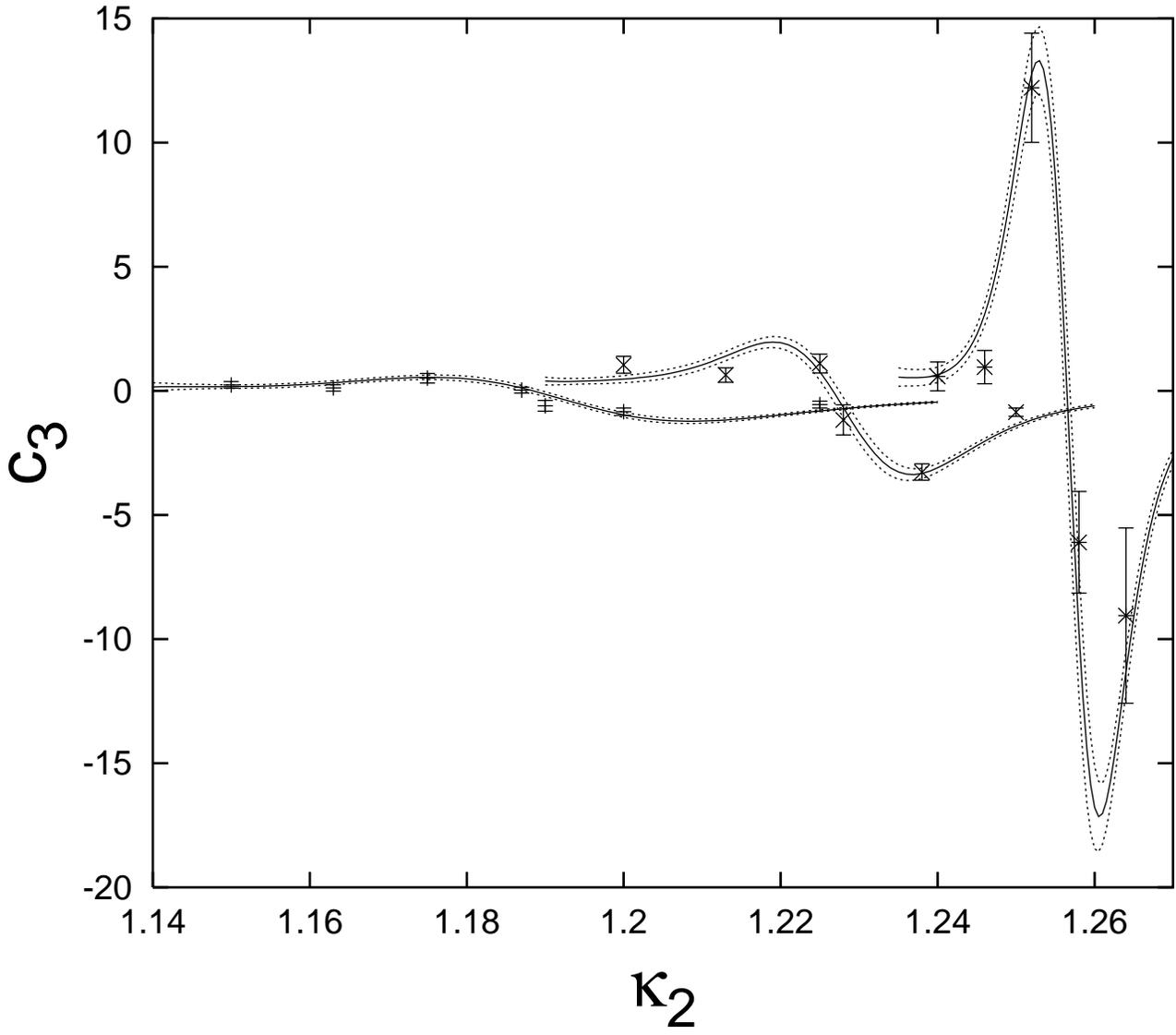
\item[Fig. 4] - The third cumulant $c_3$ as a function 
of $\kappa_2$ for lattice size 
$N_4 = 8000(+)$, $16000(\times \hspace{-0.32cm} -)$ 
and $32000(\times)$. The distance between the upper
and lower lines gives the error window.

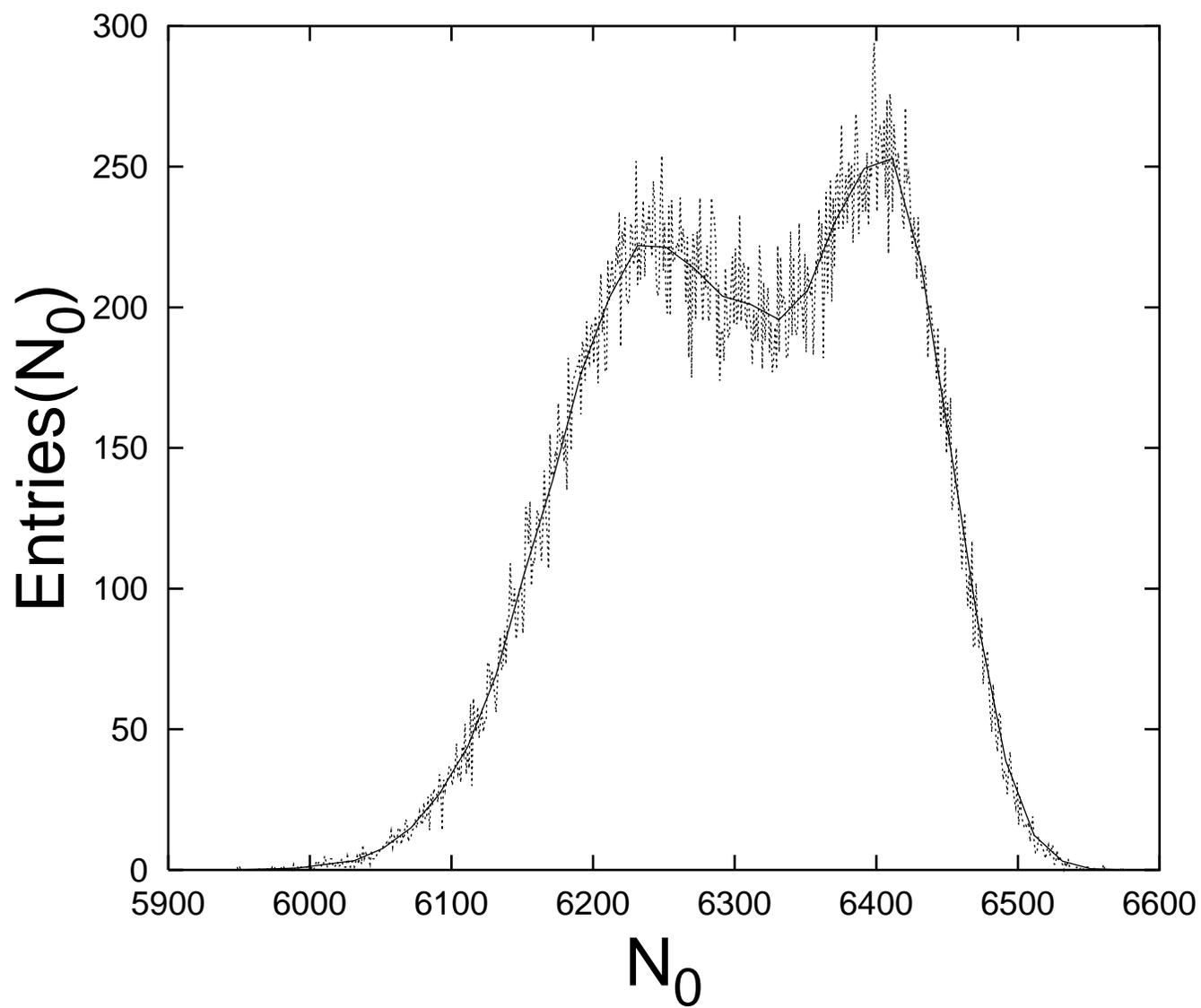
\item[Fig. 5] - The histogram for $N_0$ 
at $N_4 = 32000$ and 
$\kappa_2=1.258$.  The number of $7 \times 10^4$ 
entries gathered 
every $50$th sweep comes from $3.5 \times 10^6$ 
sweeps corresponding 
to $508$ integrated autocorrelation 
times. The solid line is 
obtained by averaging over bins of $N_0$ of length 20.

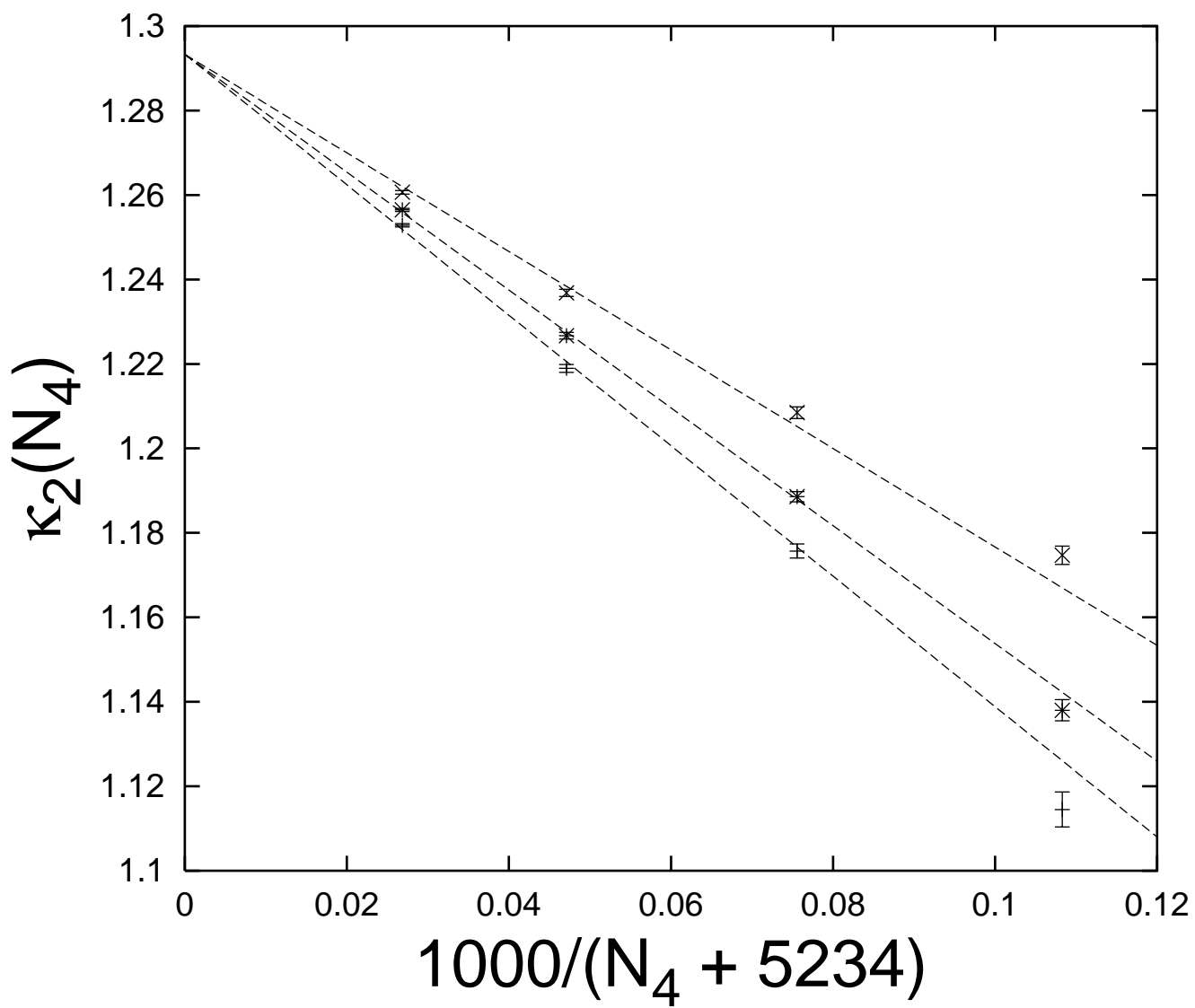
\item[Fig. 6] - The position of the maxima 
zeros and minima of the
third cumulant versus $1000/(N_4+5234)$. 
The dashed lines are obtained
fitting $\kappa_2^* + x/(N_4+a_4)$ to all those $12$ points, 
in such a way that all three lines have a common value of
$\kappa_2^*$ and $a_4$ but different values of $x$.
\end{description}

\newpage
\section*{Appendix}

\begin{table}[h]
\centering
\begin{tabular}{|r||r|r|r||r|}
\hline
&\multicolumn{3}{c||}{$N_4=4000$}
 & $N_4=3000$ \\
\hline
$\kappa_2$&$\left< r_{in}
 \right>$&$\left< r_{out}
 \right>$&$\left< N_{4out} 
\right>$&$\left< r_{in} \right>$ \\
\hline
1.025&11.93( 6)&11.03( 5)&3370(10)&-\\
1.050&12.38( 6)&11.34( 5)&3304(10)&12.09( 8)\\
1.075&12.79(11)&11.65(11)&3245(15)&12.58( 8)\\
1.100&13.59(26)&12.28(23)&3163(13)&13.36(30)\\
1.125&15.33(44)&13.88(41)&3094(10)&15.11(46)\\
1.150&16.37(62)&14.83(61)&3044(11)&16.26(32)\\
1.175&19.42(44)&17.67(43)&2965( 8)&17.52(31)\\
1.200&20.97(40)&19.20(40)&2934( 6)&18.72(28)\\
1.225&21.73(21)&19.87(21)&2910( 7)&19.35(21)\\
1.250&23.10(17)&21.19(17)&2878( 5)&-\\
\hline
\end{tabular}
\caption{RG study: data for $N_4=4000$}
\label{tab: RG4 }
\end{table}

\begin{table}[h]
\centering
\begin{tabular}{|r||r|r|r||r|}
\hline
&\multicolumn{3}{c||}{$N_4=8000$}
 & $N_4=6000$ \\
\hline
$\kappa_2$&$\left< r_{in}
 \right>$&$\left< r_{out} \right>$&$\left< N_{4out} 
\right>$&$\left< r_{in} \right>$ \\
\hline
1.100 & 14.03( 9)&12.85( 5)&
6491(22)& - \\
1.125 & 14.81(11)&13.48(11)&6327(12)&14.79(10)\\
1.150 & 15.86(17) & 14.37(17)
 & 6193(10)&16.47(34) \\
1.175&17.75(38)&16.11(37)&6055(10)&18.86(40)\\
1.200 & 23.03(49) & 21.22(48)&
5918(13) &23.19(30) \\
1.225&27.86(39)&25.94(38)&5834( 7)&25.63(28)\\
1.250 & 30.45(22) & 28.44(21)
&5783( 9)&27.44(22) \\
1.275&32.12(34)&30.05(34)&5753( 9)&28.10(16)\\
1.300 &33.96(23)&31.83(22) & 
5684( 6)&29.61(15) \\
1.325&34.46(23)&32.33(22)&5672( 6)&-\\
\hline
\end{tabular}
\caption{RG study: data for $N_4=8000$. The errors
on $N_{4out}$ are much smaller than those given in 
ref. 1. This is partly due to the use of a 
more efficient algorithm and partly to the 
elimination of a bug.}
\label{tab : RG8 }
\end{table}
\newpage
\begin{table}
\centering
\begin{tabular}{|r||r|r|r||r|}
\hline
&\multicolumn{3}{c||}{$N_4=16000$}
 & $N_4=12000$ \\
\hline
$\kappa_2$&$\left< r_{in}
 \right>$&$\left< r_{out}
 \right>$&$\left< N_{4out} 
\right>$&$\left< r_{in} \right>$ \\
\hline
1.125 & 15.42( 4)&14.22( 4)&12898(14)&-\\
1.150 & 16.20( 5) & 14.82( 4)
 & 12585(12)&16.20(11) \\
1.175&17.45(11)&15.90(11)&12284(12)& 17.19( 8)\\
1.200 & 19.08(16) & 17.38(15)&
12045( 8) &21.04(46) \\
1.213&20.85(35)&19.06(34)&11901(12)&24.92(88)\\
1.225&25.92(48)&24.01(47)&11768( 6)&28.31(92)\\
1.238&34.58(70)&32.56(70)&11655( 5)&33.18(57)\\
1.250 & 38.22(76) & 36.12(76)
&11586( 4)&35.64(39) \\
1.262&42.13(42)&40,02(42)&11529( 7)&37.62(29)\\
1.275&43.59(38)&41.41(37)&11488( 4)&38.62(27)\\
1.300 &46.01(17)&43.79(17) & 
11412( 5)&40.63(20) \\
1.325&47.88(22)&45.61(22)&11356( 5)&-\\
\hline
\end{tabular}
\caption{RG study: data for $N_4=16000$}
\label{tab : RG16 }
\end{table}

\begin{table}[p]
\centering
\begin{tabular}{|r||r|r|r|r|r|r|}
\hline
$\kappa_2$ & $c_1$ & $c_2$ & $c_3$ & $\tau$ & 
$\#$sweeps & $\# \tau$'s \\
\hline
1.075 & 0.17089(20) & 0.097(3) & 0.07( 8) & 
194(26) & 247480 & 1276 \\
1.100 & 0.17598(16) & 0.106(3) & 0.05( 6) & 
256(27) & 562090 & 2196 \\
1.125 & 0.18172(17) & 0.111(3) & 0.14( 5) & 
344(38) & 674730 & 1961 \\
1.137 & 0.18401(17) & 0.113(2) &-0.01( 5) & 
434(46) & 905340 & 2086 \\
1.150 & 0.18738(23) & 0.109(3) &-0.36( 9) & 
419(60) & 436380 & 1042 \\
1.175 & 0.19261(16) & 0.085(3) &-0.48( 7) & 
264(31) & 438020 & 1659 \\
1.200 & 0.19680(22) & 0.064(4) &-0.39(11) & 
239(42) & 163840 &  685 \\
1.225 & 0.19891(22) & 0.060(4) &-0.26( 7) & 
135(24) &  89180 &  661 \\
1.250 & 0.20170(10) & 0.049(1) &-0.16( 2) & 
 69( 7) & 163840 & 2374 \\
\hline
\end{tabular}
\caption{Cumulants: data summary for $N_4=4000$. 
We denote by $c_1$, $c_2$ and $c_3$ the first three 
cumulants of the $N_0$ distribution, normalized to the 
volume $N_4$ ($c1 = \langle N_0 \rangle/N_4$). 
The parameter $\tau$ is the integrated autocorrelation time. 
In the last two columns 
we put the number of sweeps performed and 
an estimate of the number of 
autocorrelation times to which this 
number of sweeps corresponds. }
\label{T4}
\end{table}

\begin{table}[p]
\centering
\begin{tabular}{|r||r|r|r|r|r|r|}
\hline
$\kappa_2$ & $c_1$ & $c_2$ & $c_3$ & $\tau$ & $\#$sweeps 
& $\# \tau$'s \\
\hline
1.100 & 0.16853(16) & 0.097(3) &  0.15(15) & 167( 24) & 
 163840 &  981 \\
1.125 & 0.17332(20) & 0.101(4) & -0.01(15) & 247( 45) & 
 163840 &  663 \\
1.150 & 0.17926(16) & 0.124(4) &  0.24(13) & 451( 60) & 
 572750 & 1270 \\
1.163 & 0.18237(16) & 0.117(4) &  0.12(13) & 422( 58) & 
 491520 & 1165 \\
1.175 & 0.18507(19) & 0.137(6) &  0.51(19) & 721(109) & 
 667180 &  925 \\
1.187 & 0.18859(17) & 0.143(4) &  0.03(12) & 846(113) & 
1055340 & 1247 \\
1.190 & 0.18929(22) & 0.149(5) & -0.52(19) & 939(139) & 
 694160 &  739 \\
1.200 & 0.19233(20) & 0.120(4) & -0.84(15) & 839(137) & 
 645820 &  770 \\
1.225 & 0.19762(15) & 0.069(4) & -0.55(14) & 209( 35) &  
163840 &  783 \\
1.250 & 0.20054( 7) & 0.054(2) & -0.26( 4) & 152( 14) & 
 458750 & 3018 \\
1.300 & 0.20510( 6) & 0.039(1) & -0.10( 3) &  52(  5) & 
 163840 & 3150 \\
\hline
\end{tabular}
\caption{Cumulants: data summary for 
$N_4=8000$. 
The notation is the same as for the table for $N_4=4000$.}
\label{T8}
\end{table}

\begin{table}[p]
\centering
\begin{tabular}{|r||r|r|r|r|r|r|}
\hline
$\kappa_2$ & $c_1$ & $c_2$ & $c_3$ & $\tau$ & 
$\#$sweeps & $\# \tau$'s \\
\hline
1.125 & 0.16749(21) & 0.110(9) &  0.19( 38) & 
 233( 59) &   81920 &  352 \\
1.150 & 0.17317(20) & 0.104(6) & -0.18( 29) & 
 229( 57) &   81920 &  358 \\
1.175 & 0.17914(30) & 0.130(9) &  0.13( 48) & 
 433(140) &   81920 &  189 \\
1.200 & 0.18462( 9) & 0.132(5) &  1.05( 34) & 
 647( 72) & 1262800 & 1952 \\
1.213 & 0.18813(10) & 0.133(4) &  0.64( 29) & 
 726( 89) & 1136040 & 1565 \\
1.225 & 0.19228(18) & 0.190(7) &  1.10( 38) & 
2022(332) & 1551360 &  767 \\
1.228 & 0.19388(23) & 0.198(6) & -1.17( 61) & 
2342(476) & 1396500 &  596 \\
1.238 & 0.19669(15) & 0.144(6) & -3.27( 33) & 
1735(274) & 1419640 &  818 \\
1.250 & 0.19928(16) & 0.083(4) & -0.86( 17) & 
 596(137) &  245760 &  412 \\
1.262 & 0.20113(13) & 0.054(4) & -0.59( 21) & 
 174( 39) &   81920 &  471 \\
1.275 & 0.20241( 7) & 0.049(3) & -0.28( 13) & 
 108( 15) &  163840 & 1517 \\
1.300 & 0.20454( 5) & 0.040(1) & -0.12(  6) &   
50(  5) &  163840 & 3276 \\
1.325 & 0.20644( 5) & 0.037(1) & -0.09(  5) & 
  36(  4) &  122880 & 3413 \\
\hline 
\end{tabular}
\caption{Cumulants: data summary for 
$N_4=16000$. 
The notation is the same as for the table for $N_4=4000$.}
\label{T16}
\end{table}

\begin{table}[p]
\centering
\begin{tabular}{|r||r|r|r|r|r|r|}
\hline
$\kappa_2$ & $c_1$ & $c_2$ & $c_3$ & $\tau$ & 
$\#$sweeps & $\# \tau$'s \\
\hline
1.240 & 0.18970(12) & 0.141( 7) &  0.58( 58) & 
1046( 188) &  674900 &645 \\
1.246 & 0.19150(11) & 0.144( 8) &  0.96( 67) & 
1124( 194) &  786950 &700 \\
1.252 & 0.19399(32) & 0.254(35) & 12.21(220) & 
6419(2183) & 1000350 &156 \\
1.258 & 0.19712(20) & 0.316( 8) & -6.10(205) & 
6883(1352) & 3502350 &508 \\
1.264 & 0.20052(21) & 0.118(20) & -9.05(353) & 
4064(1326) &  695650 &171 \\
1.270 & 0.20085(27) & 0.118(20) & -6.56(163) & 
3027(1269) &  304850 &101 \\
\hline
\end{tabular}
\label{T32}
\caption{Cumulants: data summary for 
$N_4=32000$. The notation is the same 
as for the table for $N_4=4000$.}
\end{table}

\begin{table}[p]
\centering
\begin{tabular}{|r||r|r|r||r|r|r|}
\hline
$N_4$ & $c_{2 \, max}$ & $c_{3 \, max}$ & $c_{3 \, min}$ & 
$\kappa_{2 \, max}$ & $\kappa_{3 \, max}$ 
& $\kappa_{3 \, min}$ \\
\hline
 4K& 0.116( 1)& 0.18( 5)& -0.53( 3)& 1.1380(25)& 
1.1145(41)& 1.1747(22) \\
 8K& 0.143( 3)& 0.54( 9)& -1.21( 9)& 1.1886(12)& 
1.1756(17)& 1.2085(14) \\
16K& 0.191( 5)& 1.94(22)& -3.35(24)& 1.2267( 8)& 
1.2189( 9)& 1.2368( 8) \\
32K& 0.335(13)& 13.2(13)& -17.0(14)& 1.2565( 4)& 
1.2528( 4)& 1.2606( 4) \\
\hline
\end{tabular}
\label{Textrema}
\caption{The heights and positions of the 
maximum of the second cumulant
and of the maximum and minimum of the third 
one for different volumes.}
\end{table}

\end{document}